\documentclass{tran-l}

\theoremstyle{definition}

\theoremstyle{remark}

\numberwithin{equation}{section}

\begin{document}

\title[Energy Distribution of a  Stationary Beam of Light]
{Energy Distribution of a Stationary Beam of Light}

\author{ Ragab M. Gad}
\address{ Mathematics Department, Faculty of
Science, Minia University, El-Minia, Egypt}

\email{ragab2gad@hotmail.com}






\dedicatory{}


\large{
\begin{abstract}

Aguirregabiria et al showed that Einstein, Landau and Lifshitz, Papapetrou,
and Weinberg energy-momentum complexes coincide for all Kerr-Schild metric.
Bringely used their general expression of the Kerr-Schild class and found 
energy
and  momentum densities for the Bonnor metric. We obtain these results 
without using
Aguirregabiria  et al results and verify that Bringley's results are 
correct. This also supports
Aguirregabiria  et al results as well as Cooperstock hypothesis. Further, we 
obtain the energy
distribution of the space-time under consideration
\end{abstract}

\maketitle
\section{Introduction}

One of the most interesting problems which remains unsolved since
Einstein proposal of general theory of relativity, is the energy-momentum
localization. After Einstein \cite{1} obtained an expression for the
energy-momentum complexes many physicists, such as Landau and
Lifshtz \cite{2}, Tolman \cite{3}, Papapetrou \cite{4}, and Weinberg 
\cite{5} had
given different definitions for the energy-momentum complex. These
definitions were restricted to evaluate energy disribution in 
quasi-cartesian
coordinates. This motivated M{\o}ller \cite{6} and many other, like Komar 
\cite{7}
and Penrose \cite{8}, to construct coordinate independent definitions.
Each of these has its own drawbacks (see \cite{9}).
\par
In the past decade much work has been produced showing the Einstein,
Tolman, Landau and Lifshitz, Papapetrou and Weinberg (later ETLLPW)
complexes give meaningful results for many well known space-times
(see \cite{10}-\cite{22}).
Aguirregabiria et al \cite{23} showed that the energy-momentum complexes
of ETLLPW give the same result for any Kerr-Schild (KS) metric if the
computations are performed in Kerr-Schild cartesian coordinates.\\
The energy and momentum densities are calculated for a stationary
beam of light using the definitions of ELLPW. It is shown that the four
energy-momentum complexes of ELLPW concide if the calculations
are carried out in Kerr-Shild cartesian coordinates.
\par
In the present paper we calculate the five energy-momentum complexes
of ETLLPW for the Bonnor metric if the calculations are performed in
cartesian coordinates.
\par
We use the convention that Latin indices take values from 0 to 3 and Greek
indices take values from 1 to 3, and take units where $G = 1$ and $c = 1$.
\section{The Bonnor Metric}
The Bonnor metric is given by \cite{24}
\begin{equation}\label{BM}
ds^2 = (1 + m) dt^2 - 2mdt dz - dx^2 - dy^2 - (1- m)dz^2,
\end{equation}
where $m$ is a function of $x$ and $y$ only.\\
The non-zero components of the energy momentum tensor $T^{a}_{b}$
$$
- T^3_3 = - T^0_3 = T^3_0 = T^0_0 = \rho,
$$
where $\rho$ is the energy density, and
\begin{equation}\label{cond.}
\nabla^2 m = 16 \pi \rho .
\end{equation}
$m$ must be such $\rho$ is non-negative.
The line element (\ref{BM}) describing a stationary
beam of light flowing in the z-direction.
\par
The determinant and the non-zero components of the contravariant metric
tensor are
\begin{equation}\label{2.3}
\begin{split}
g & = -1\\
g^{00} & = 1 - m, \\
g^{03} & = - m,\\
g^{11} & =  - 1,\\
g^{22} & = - 1,\\
g^{33} & = - (1+ m).
\end{split}
\end{equation}
Now, we can require the following list of non-vanishing components
of the Christoffel symbol
\begin{align*}
\Gamma_{01}^{0} & = \frac{1}{2}m_{x}, &
\Gamma_{02}^{0} & = \frac{1}{2}m_{y},\\
\Gamma_{13}^{0} & = - \frac{1}{2}m_{x},&
\Gamma_{23}^{0} & = - \frac{1}{2}m_{y},\\
\Gamma_{00}^{1} & =  \frac{1}{2}m_{x},&
\Gamma_{03}^{1} & = - \frac{1}{2}m_{x},\\
\Gamma_{33}^{1} & =  \frac{1}{2}m_{x},&
\Gamma_{00}^{2} & =  \frac{1}{2}m_{y},\\
\Gamma_{03}^{2} & = - \frac{1}{2}m_{y},&
\Gamma_{33}^{2} & =  \frac{1}{2}m_{y},\\
\Gamma_{01}^{3} & =  \frac{1}{2}m_{x},&
\Gamma_{02}^{3} & =  \frac{1}{2}m_{y},\\
\Gamma_{13}^{3} & = - \frac{1}{2}m_{x},&
\Gamma_{23}^{3} & =  \frac{1}{2}m_{y}.
\end{align*}
where subscripts denote partial differentiation.

\section{Energy-momentum Complex as Defined by Einstein}
The energy-momentum complex as defined by Einstein is given by
\begin{equation} \label{3.1}
\theta^{k}_{i} = \frac{1}{16\pi}H^{kl}_{\, \,\, i,l},
\end{equation}
where
\begin{equation} \label{3.2}
H^{kl}_{\, \,\, i} = - H^{lk}_{\, \,\, i} = \frac{g_{in}}{\sqrt{- g}}\big[ - 
g\big( g^{kn}g^{lm} - g^{ln}g^{km}\big)\big]_{,m}.
\end{equation}
$\theta^{0}_{0}$ and $\theta^{0}_{\alpha}$ are the energy and momentum
density components, respectively.\\
The energy-momentum complex $\theta^k_i$ satisfies the local conservation
law
$$
\frac{\partial\theta^k_i}{\partial x^k} = 0
$$
\par
The energy and momentum in Einstein's prescription are given by
\begin{equation}\label{3.3}
P_{i} = \int\int\int \theta^{0}_{i} dx^1 dx^2 dx^3.
\end{equation}
Using the Gauss's theorem, we get
\begin{equation}\label{3.4}
P_{i} = \frac{1}{16\pi}\int\int {H^{0\alpha}_{\, \,\, i} n_{\alpha}} ds,
\end{equation}
where $n_{\beta}$ is the outward unit normal vector over an
infinitesimal surface element $ds$. $P_{0}$ and $P_{\alpha}$ are
the energy and momentum components.\\
In order to evaluate the energy and momentum densities in Einstein's
prescription associated with the Bonnor space-time, we evaluate the
non-zero components of $H^{kl}_{\, \,\, i}$

\begin{equation}\label{3.5}
\begin{split}
H^{01}_{\, \,\, 0} & = m_x,\\
H^{02}_{\, \,\, 0}  & = m_y, \\
H^{01}_{\, \,\, 3} & = - m_x,\\
H^{02}_{\, \,\, 3}  & = - m_y.
\end{split}
\end{equation}
Using these components and equation (\ref{cond.}) in equation (\ref{3.1}),
we get the energy and momentum densities as following
\begin{equation}\label{3.6}
\theta^{0}_{0} = \rho
\end{equation}
\begin{equation}
\theta^{03} = \eta^{33}\theta^{0}_{3} = \rho.
\end{equation}
Using (\ref{3.6}) in equation (\ref{3.3}), we get the expression of the
energy in the form
\begin{equation}
E_{Ein} = M.
\end{equation}
\section{Energy Distribution in Tolman's Prescription}
The energy-momentum complex of Tolman  is
\begin{equation}\label{4.1}
\Im_{k}^{i}
= \frac{1}{8\pi }U^{ij}_{k,j},
\end{equation}
where

$$
U^{ij}_{k} =
\sqrt{-g}\big[ -g^{pi}\big(-\Gamma^{j}_{kp} + \frac{1}{2}\delta
^{j}_{k}\Gamma^{a}_{ap} +
\frac{1}{2}\delta^{j}_{p}\Gamma^{a}_{ak}\big)
$$
\begin{equation}\label{4.2}
+ \frac{1}{2}\delta^{i}_{k}g^{pm}\big(-\Gamma^{j}_{pm} +
\frac{1}{2}\delta^{j}_{p}\Gamma^{a}_{am} +
\frac{1}{2}\delta^{j}_{m}\Gamma^{a}_{ap}\big)\big],
\end{equation}
$\Im_{0}^{0}$ is the energy density, $\Im^{\alpha}_{0}$ are the
components of the energy current density, and $\Im^{0}_{\alpha}$
are the momentum density components. \\
The energy-momentum
complex $\Im_{k}^{i}$ satisfies the local conservation law
$$
\frac{\partial \Im_{k}^{i}}{\partial x^{i}} = 0.
$$
The energy
distribution in the Tolman' definition $E_{Tol}$ is given by
\begin{equation}\label{4.3}
E_{Tol} =
\int\int\int{\Im^{0}_{0}} dx dy dz.
\end{equation}
Using the Gauss theorem
(noting that the space-time under consideration is static), one has
\begin{equation}\label{4.4}
E_{Tol} = \frac{1}{8\pi}\int\int{U^{0\beta}_{0}
n^{(\alpha)}_{\beta}} ds_{(\alpha)},
\end{equation}
where $n_{\beta}$ is the
unit vector over an infinitesimal surface element, $ds$.
\par
Using (\ref{2.3}) and the components of Christoffel symbol in (\ref{4.2}), 
after
straightforward but rather lengthy calculations, we get
\begin{equation}\label{4.5}
\begin{split}
U^{01}_{0} & =  \frac{1}{2}m_{x},\\
U^{02}_{0} & =  \frac{1}{2}m_y,\\
U^{01}_{3} & = - \frac{1}{2}m_{x},\\
U^{02}_{3} & = - \frac{1}{2}m_y,\\
\end{split}
\end{equation}
Using (\ref{4.5}) and
the condition (\ref{cond.}) in (\ref{4.1}), we obtain the
energy density and momentum density in the form
\begin{equation}\label{4.6}
\Im_{0}^{0} = \rho
\end{equation}
\begin{equation}
\Im^{03} = \eta^{33}\Im^{0}_{3} = \rho.
\end{equation}
Using (\ref{4.6}) in equation (\ref{4.3}), then the energy distribution
associated with the Bonnor space-time is given by
\begin{equation}\label{4.7}
E_{Tol}  = M.
\end{equation}

\section{The Landau and Lifshitz Energy-momentum Complex}
Landau and Lifshitz's energy-momentum complex  is given by
\begin{equation}\label{5.1}
L^{ij} = \frac{1}{16\pi}S^{ikjl}_{\quad ,kl},
\end{equation}
where
\begin{equation}\label{5.2}
S^{ikjl} = -g\big( g^{ij}g^{kl} - g^{il}g^{kj}\big).
\end{equation}
$L^{ij}$ is symmetric in its indices, $L^{00}$ is the energy density and
$L^{0\alpha}$ are the momentum (energy current) density components.
$S^{ikjl}$ has symmetries of the Riemann curvature tensor. \\
The expression
\begin{equation}\label{5.3}
P^i  = \int\int\int{L^{i0}}dx^1 dx^2 dx^3
\end{equation}
gives the energy $P_0$ and the momentum $P_{\alpha}$  components.
Further Gauss's theorem furnishes the energy
$E_{LL}$ given by
\begin{equation}\label{5.4}
E_{LL} =
\frac{1}{16\pi}\int\int{S^{0\alpha 0\beta}_{\quad \alpha}n_{\beta}}ds,
\end{equation}
where $n_{\beta}$ is the outward unit normal vector over an infinitesimal
surface element $ds$.\\

The non-zero components of $S^{ikjl}$ for the line element (\ref{BM}) are
\begin{equation}\label{5.5}
\begin{split}
S^{0101} & = - (1 - m),\\
S^{0202} & = - (1 - m),\\
S^{0311} & = m,\\
S^{0322} & = m.
\end{split}
\end{equation}
Now, substituting (\ref{5.5})  and using the condition (\ref{cond.}), we 
obtain the
energy and momentum densities of the line element (\ref{BM}) in the sense of
Landau and Lifshitz in the following form \begin{equation}\label{5.6}
L^{00} = \rho.
\end{equation}
\begin{equation}
L^{03} = \rho
\end{equation}

Using (\ref{5.6})  in equation (\ref{5.3}), we get the energy
\begin{equation}
E_{LL} = M.
\end{equation}
\section{The Energy-Momentum Complex of Papapetrou}
The symmetric energy-momentum complex of Papapetrou  is given by
\begin{equation}\label{6.1}
\Omega^{ij} = \frac{1}{16\pi} \Upsilon^{ijkl}_{\quad ,kl},
\end{equation}
where
\begin{equation}\label{6.2}
\Upsilon^{ijkl} = \sqrt{- g}\big( g^{ij}\eta^{kl} - g^{ik}\eta^{jl} + 
g^{kl}\eta^{ij} - g^{jl}\eta^{ik}\big),
\end{equation}
and $\eta^{ik}$ is the Minkowski metric with signature $-2$.\\
$\Omega^{00}$ and $\Omega^{0\alpha}$ are the energy and momentum
density components. The energy and momentum components are given by
\begin{equation}\label{6.3}
P^i = \int\int\int{\Omega^{i0}} dx^1 dx^2 dx^3.
\end{equation}
Using the Gauss theorem, The energy $E_{P}$ takes the following form
\begin{equation}\label{6.4}
E_{P} = \frac{1}{16\pi}\int\int{\Upsilon^{00\alpha\beta}_{\quad 
,\beta}n_{\alpha}}ds.
\end{equation}
To find the energy and momentum densities of the space-time under 
consideration, we require the
following non-zero components of $\Upsilon^{ijkl}$
\begin{equation}\label{6.5}
\begin{split}
\Upsilon^{0011} & = m - 2,\\
\Upsilon^{0022} & = m - 2,\\
\Upsilon^{0311} & = m,\\
\Upsilon^{0322} & = m.
\end{split}
\end{equation}
Using these components in (\ref{6.1}), we get the energy and momentum 
densities in the following form
\begin{equation}\label{6.6}
\Omega^{00} = \rho.
\end{equation}
\begin{equation}
\Omega^{03} = \rho.
\end{equation}
The energy $E_{P}$ of the Bonnor space-time is obtained by using (\ref{6.6}) 
in (\ref{6.3})
\begin{equation}
E_{P} = M.
\end{equation}

\section{The Weinberg Energy-Momentum Complex}
The symmetric energy-momentum complex of Weinberg  is given by
\begin{equation}\label{7.1}
W^{ij} = \frac{1}{16\pi}\Delta^{ijk}_{\quad ,k},
\end{equation}
where
$$
\Delta^{ijk} = \frac{\partial h^{a}_{a}}{\partial x_{I}}\eta^{jk} -
\frac{\partial h^{a}_{a}}{\partial x_{j}}\eta^{ik} - \frac{h^{ai}}{\partial 
x^A}\eta^{jk} +
\frac{\partial h^{aj}}{\partial x^a}\eta^{ik}
$$
\begin{equation}\label{7.2}
+ \frac{\partial h^{ik}}{\partial x_{j}} - \frac{\partial h^{jk}}{\partial 
x_{i}},
\end{equation}
and
$$
h_{ij} = g_{ij} - \eta_{ij}.
$$
$\eta_{ij}$ is the Minkowski metric with signature $- 2$. The indices on 
$h_{ij}$ or $\frac{\partial}{\partial x_{i}}$ are raised or lowered with the 
help of $\eta$'s. The Weinberg energy-momentum complex $W^{ij}$ satisfies 
the local conservation laws
$$
\frac{\partial W^{ij}}{\partial x^{j}} = 0.
$$
$W^{00}$ and $W^{\alpha 0}$ are the energy and momentum density components. 
The energy and momentum components are given by
\begin{equation}\label{7.3}
P^{i} = \int\int{W^{i0}} dx^1 dx^2 dx^3.
\end{equation}
The only required components of $\triangle^{ijk}$ in the calculation of the 
energy and momentum densities are the following
\begin{equation}\label{7.4}
\begin{split}
\triangle^{001} & = m_{x},\\
\triangle^{002} & = m_{y},\\
\triangle^{031} & = m_{x},\\
\triangle^{032} & = m_{y}.
\end{split}
\end{equation}
Using these results in (\ref{7.1}) and using the condition (\ref{cond.}), we 
obtain the
energy density and momentum density of the space-time (\ref{BM})
\begin{equation}\label{7.5}
W^{00} = \rho.
\end{equation}
\begin{equation}
W^{03} = \rho.
\end{equation}

Using (\ref{7.5}) in  (\ref{7.3}) the energy $E_{W}$ is
\begin{equation}
E_{W} = M.
\end{equation}
\section{Summary}

It is well-known that the subject of the energy-momentum
localization is associated with much debate. Misner et al. \cite{MTW73}
argued that the energy is localizable only for spherical systems.
Cooperstock and Sarracino \cite{CS78} gave their viewpoint that if the
energy is localizable in spherical systems then it is localizable
for all systems. Bondi \cite{B90} gave his opinion that a
nonlocalizable form of energy is not admissible in relativity so
its location can in principle be found.\\
Some interesting results which have been found recently (see for example, 
\cite{10}, \cite{15}, \cite{16}
) lend support to the idea that the several energy-momentum
complexes can give the same and acceptable result for a given
space-time. Virbhadra \cite{V99} emphasized that though that the
energy-momentum complexes are non-tensors under general coordinate
transformations, the local conservation laws with them hold in all
coordinate system.  Aguirregabiria et al \cite{23} showed that
different energy-momentum complexes yield tha same energy
distribution for any Kerr-Schild class metric.
\par
In this paper we obtained the energy distribution associated
with the Bonnor metric describing a stationary beam of light.
We used the energy-momentum complexes of Einstein,
Tolman, Landau and Lifshitz, Papapetrou and Weinberg.
All these definitions give the same results for the energy density, momentum
density, and energy distribuation.
Our calculations are performed in Cartesian coordinates.\\
The results for the energy and momentum densities obtained here are the same 
as the results
obtained by Bringley \cite{21} using the energy-momentum
complexes of ELLPW in Kerr-Schild cartesian coordinates.\\
Our results sustain the Aguirregabiria  et al. \cite{23} results as well as
Cooperstock hypothesis \cite{C99} (which essentially states that the
energy and momentum in a curved space-time are confined to the regions
of non-vanishing energy-momentum tensor $T^{i}_{j}$ of the matter and all
non-gravitational fields).

\section*{Acknowledgments}
I am grateful to Professor K. S. Virbhadra for his helpful advice.

}
\end{document}